\documentclass[prl,twocolumn,nofootinbib]{revtex4-1}
\usepackage{graphicx}
\usepackage{color}
\usepackage{dcolumn}
\usepackage{bm}
\usepackage{amssymb}
\usepackage[bottom]{footmisc}
\usepackage{footnote}
\usepackage{lipsum}
\usepackage{wrapfig}
\usepackage{sidecap}
\renewcommand{\thefootnote}{\fnsymbol{footnote}}

\newcommand\blfootnote[1]{%
\begingroup
\renewcommand\thefootnote{}\footnote{#1}%
\addtocounter{footnote}{-1}%
\endgroup}
\begin{document}
\title{Electron optics with ballistic graphene junctions}
\author{ Shaowen Chen,$^{1,2\ast}$ Zheng Han,$^{1,7\ast }$ Mirza M. Elahi,$^{3}$ K. M. Masum Habib,$^{3\dagger}$ Lei Wang,$^{4}$ Bo Wen,$^{1}$ Yuanda Gao,$^{5}$ Takashi Taniguchi,$^{6}$ Kenji Watanabe,$^{6}$ James Hone,$^{5}$ Avik W. Ghosh,$^{3}$ and Cory R. Dean$^{1\ddagger}$}
\affiliation{$^{1}$Department of Physics, Columbia University, New York, NY 10027, USA}
\affiliation{$^{2}$Department of Applied Physics and Applied Mathematics, Columbia University, New York, NY 10027, USA}
\affiliation{$^{3}$Department of Electrical and Computer Engineering, University of Virginia, Charlottesville, VA 22904, USA}
\affiliation{$^{4}$Department of Physics, Cornell University, Ithaca , NY 14853, USA}
\affiliation{$^{5}$Department of Mechanical Engineering, Columbia University, New York, NY 10027, USA}
\affiliation{$^{6}$National Institute for Materials Science, 1-1 Namiki, Tsukuba 305-0047, Japan}
\affiliation{$^{7}$Shenyang National Laboratory for Materials Science, Institute of Metal Research, Chinese Academy of Sciences, Shenyang 110016, P. R. China}
\date{\today}
\maketitle
\blfootnote{\textup{*} These authors contribute equally.}
\blfootnote{$^\dagger$Current address: Intel Corp., Santa Clara  CA 95054, USA}
\blfootnote{$^\ddagger$E-mail:  cdean@phys.columbia.edu}

\textbf{Electrons transmitted across a ballistic semiconductor junction undergo refraction, analogous to light rays across an optical boundary. A $pn$ junction theoretically provides the equivalent of a negative index medium, enabling novel electron optics such as negative refraction and perfect (Veselago) lensing. In graphene, the linear dispersion and zero-gap bandstructure admit highly transparent $pn$ junctions by simple electrostatic gating, which cannot be achieved in conventional semiconductors. Moreover ballistic transport over micron length scales at ambient temperature has been realized, providing an ideal platform to realize a new generation of device based on electron lensing. Robust demonstration of these effects, however, has not been forthcoming. Here we employ transverse magnetic focusing to probe propagation across an electrostatically defined graphene junction. We find perfect agreement with the predicted Snell'€™s law for electrons, including observation of both positive and negative refraction. Resonant transmission across the $pn$ junction provides a direct measurement of the angle dependent transmission coefficient, and we demonstrate good agreement with theory.  Comparing experimental data with simulation reveals the crucial role played by the effective junction width, providing guidance for future device design. Our results pave the way for realizing novel electron optics based on graphene $pn$ junctions.} 

Ballistic electrons in a uniform 2D electron gas (2DEG) behave in close analogy to light\cite{Houten1995,Dragoman_ProgQuantElec1999}: electrons follow straight-line trajectories and their wave nature can manifest in a variety of interference and diffraction effects. When transmitted across a boundary separating regions of different density, electrons undergo refraction\cite{Spector_APL,Shtrikman_PRB_ELens}, much like light rays crossing a boundary between materials with different optical index. This makes it possible to manipulate electrons like photons, using optics-inspired components such as mirrors, lenses, prisms and splitters\cite{Shtrikman_PRB_ELens,Spector_APL,Cheianov_Science,Park_NanoLett2008}. A particularly striking feature of electronic optics is the prediction of negative refraction\cite{Veselago_SovPhysUsp1968}, which is difficult to achieve in photonic systems but conceptually straightforward for electrons, arising when carriers cross a $pn$ junction separating electron and hole bands.  In optical metamaterials\cite{Shalaev_NatPhot2007,Pendry_PerfectLens, shelby2001experimental}, negative refraction is enabling exotic new device technologies such as superlenses\cite{fang2005sub}, which can focus beyond the diffraction limit, and optical cloaks\cite{schurig2006metamaterial}, which make objects invisible by bending light around them. 

\begin{figure}[ht]
\includegraphics[width=1\linewidth]{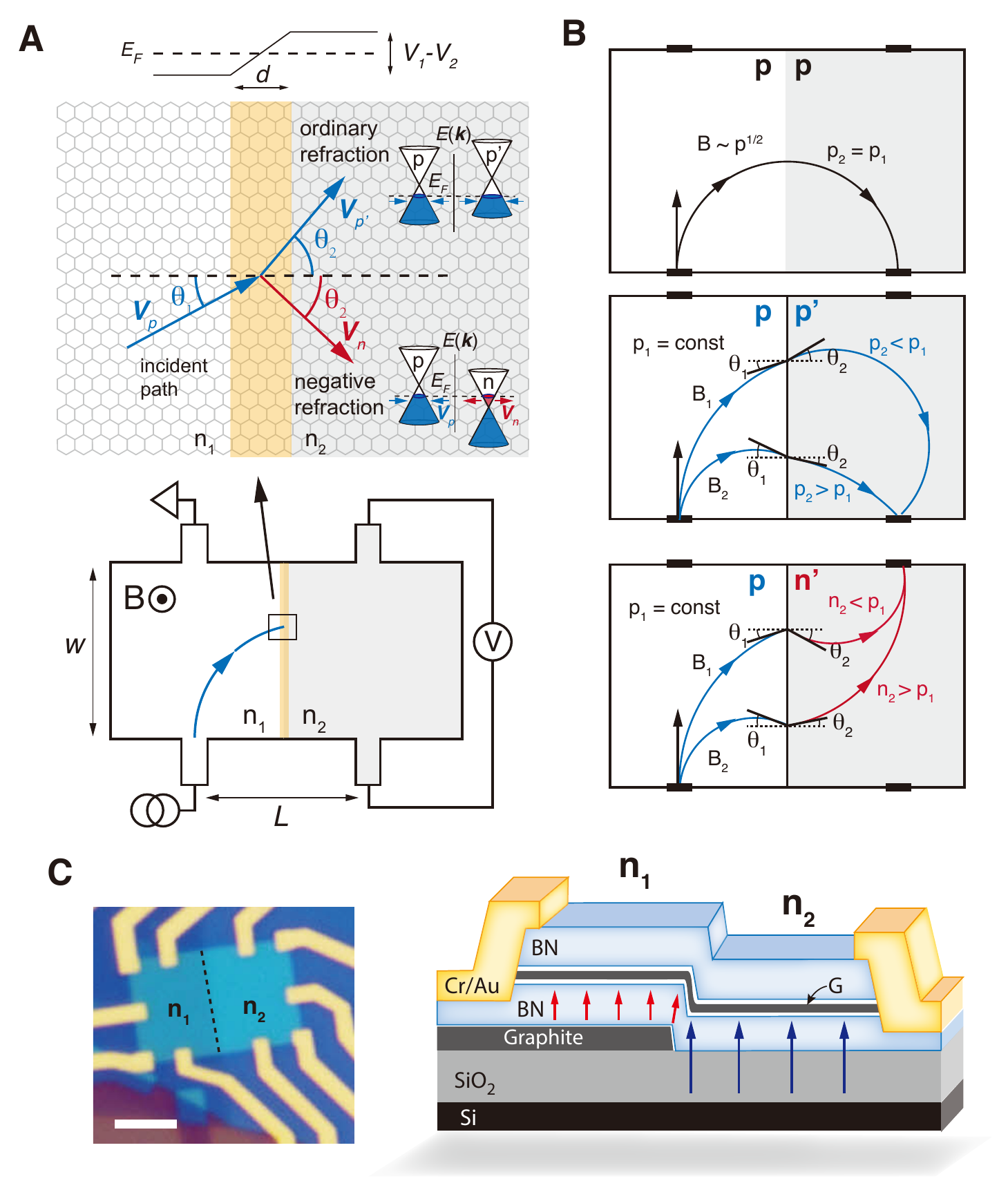} 
\caption{\textbf{Electron Refraction.} (\textbf{A}) Transverse magnetic field is used to focus electrons onto a split gate junction at variable incident angles. The cyclotron radius, determined by the magnetic field and Fermi momentum (or related carrier density), determines the incidence angle. The density difference across the boundary, induced by the two gate voltages, determines the refraction angle (see text). (\textbf{B}) A resonant path is shown for three example scenarios corresponding to $pp$ (equal hole density), $pp'$ (unequal hole density) and $pn'$ (unequal hole-electron densities). In our measurement scheme, density $n_{1}$ is fixed, while varying B and $n_{2}$. (\textbf{C}) Optical image (left) and cartoon schematic (right) of split gate device. A naturally cleaved graphite edge is utilized to define an atomically smooth electrostatic boundary. Scale bar is 5 $\mu$m.}
\end{figure}

Graphene has been considered an ideal platform for demonstrating electron optics in the solid state \cite{Cheianov_Science,Internal_total_reflection,rickhaus2013ballistic,taychatanapat2015conductance}. The high intrinsic mobility allows ballistic transport over micrometer length scales at ambient temperatures \cite{Dean}, while the lack of a bandgap makes graphene $pn$ junctions highly transparent\cite{Cheianov_Science,CheinovPRB2006,huard2007transport,gorbachev2008conductance,young2009quantum,stander2009evidence,RedwanPRB2012,grushina2013ballistic,rickhaus2013ballistic,rickhaus2015snake} compared with conventional semiconductors. However, experimental demonstration of electron lensing in graphene junctions, has remained conspicuously difficult to realize:  separating the junction response from mesoscopic effects (such as contacts and boundary scattering) in transport experiments has proven difficult, while direct probe techniques \cite{Topinka_Nature_2001,Westervelt_NatPhys_2007,BaringhausSciReport2015} have not provided real-space mapping of transmission across a junction. Here we demonstrate that by utilizing a transverse magnetic focusing (TMF) measurement scheme in a split gate device, we are able to isolate and measure directly the relationship between the incident and refracted electron trajectories. We confirm an electronic Snell's law relation and find unambiguous evidence of negative refraction across a $pn$ junction. Our technique additionally provides a direct quantitative measure of the transmission coefficient with incidence angle, which we find to be in excellent agreement with theory\cite{CheinovPRB2006,RedwanPRB2012}. Together with semi-classical simulations, our results reveal the crucial role played by the junction profile for electron optics, and provide a roadmap for new device technologies based on graphene $pn$ junctions.

\begin{figure*}[ht]
\includegraphics[width=.95\linewidth]{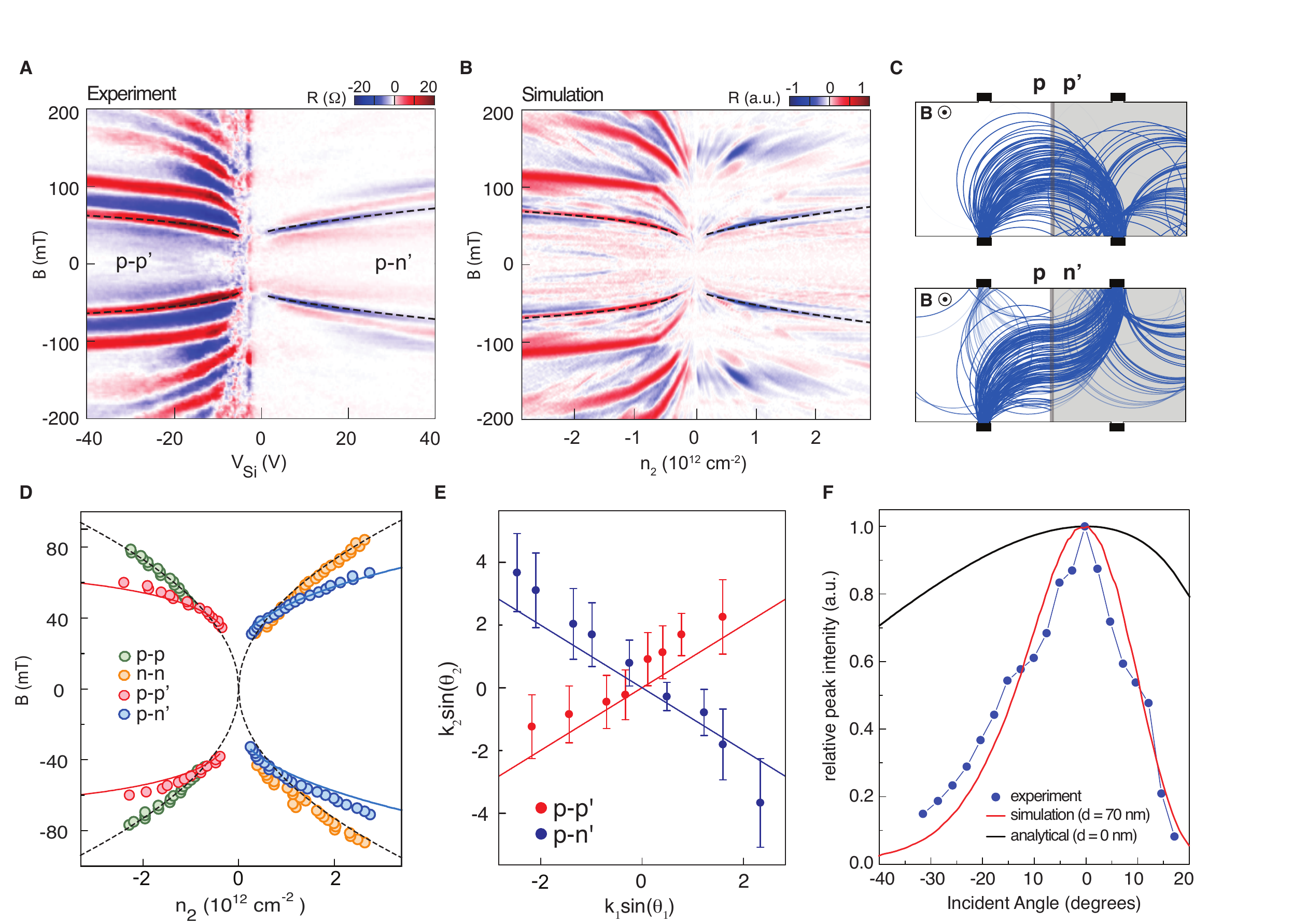}
\caption{\textbf{Snell's law for electrons.} (\textbf{A}) Resistance parallel to the junction (corresponding to the measurement configuration shown in Fig. 1b) versus magnetic field and Silicon gate, $V_{Si}$. The graphite gate region is fixed to constant $p$-type carrier density ($V_{G}=-1$~V). (\textbf{B}) Simulation of the experimental data in (A), from  ray tracing paths.  Representative resonant electron trajectories are shown in (\textbf{C}) for a $pp'$ (top) and $pn'$ (bottom) junction.  (\textbf{D}) Position of the points plotted as $B$ versus $n_{2}$ from the lowest order resonance modes.  $pp'$ and $pn'$ data points are taken from (A).  $pp$ and $nn$ data points are determined from a similar map in which the gates are synchronized to maintain a matched density (see SI).  Dashed line represents the theoretical resonance condition for graphene with matched density (i.e. no junction). Solid red line and blue lines are the theoretical curves deduced from our geometric model, including refraction, for $pp'$ and $pn'$ junctions, respectively (see text). (\textbf{E}) Snell's law parameters calculated from the peak points (see text). (\textbf{F}) Transmission intensity versus incident angle.  Blue circles correspond to the normalized peak resistance values extracted from (A). Red line is the normalized intensity from simulation for a device with a graded junction of width 70~nm.  Black line is the theoretical angle dependance for an abrupt ($d=0$~nm) junction.}
\label{fig:fig2}
\end{figure*}

\begin{figure}
\includegraphics[width=0.95\linewidth]{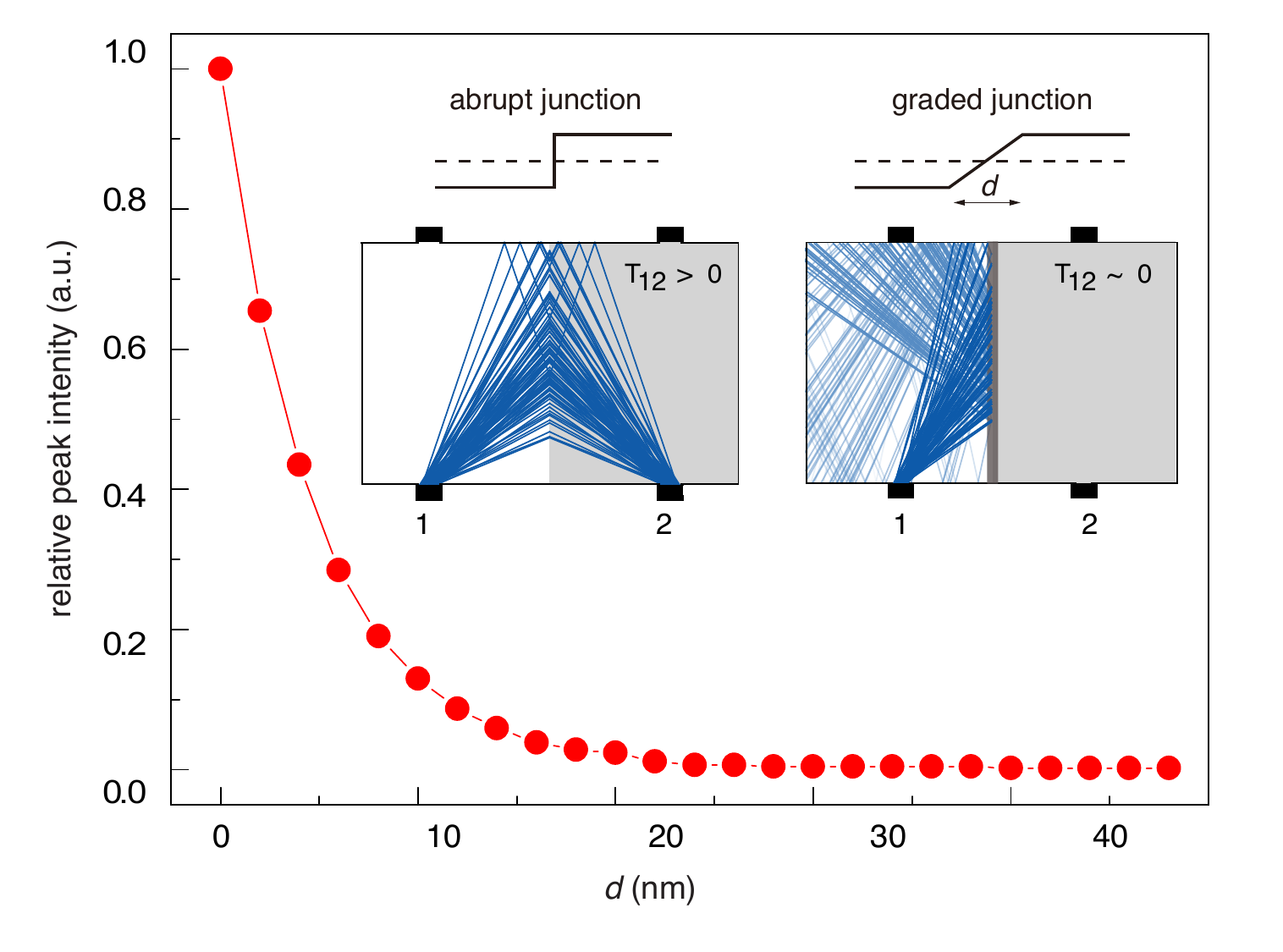}
\caption{\textbf{Veselago lensing.} Transmission coefficient for electrons focused across a $pn$ junction.  Main panel shows the variation in transmission probability versus junction width $d$, determined from simulation.  Diverging electrons across a $pn$ junction theoretically converge to an equidistant point owing to negative refraction.  For a graded junction the majority of the electrons are reflected, explaining why Veselago focusing is not observed. Inset shows representative simulated electron trajectories for an abrupt (left) and graded (right) junction.}
\label{fig:fig3}
\end{figure}

For electrons, conservation of the transverse component of the momentum vector, $\textbf{k}$, across the junction leads to the Snell's law relation $k_1\sin{\theta_1}=k_2\sin{\theta_{2}}$, where $\theta_1$ and $\theta_2$ are the incident and refracted angle with respect the boundary normal, and the Fermi wavevector, $k_i=\sqrt{\pi n_i}$ replaces the optical index of refraction. Since the group velocity is defined by the energy band dispersion $\textbf{v}=dE/d(\hbar k)$, the sign changes between the valence and conduction bands, making it parallel to the Fermi momentum for $n-$type carriers, but antiparallel for $p-$type. In the case of a $pn$ junction, the transverse component of the group velocity must change sign in order to conserve momentum (Fig. 1a) and a negative refraction angle results. 

Fig. 1a-b illustrates the device structure used to test this relation.  A sample with a junction separating areas of different carrier density is contacted by multiple electrodes in both regions. Under a transverse magnetic field, injected electrons undergo cyclotron motion with radius determined by the Lorentz force. In the absence of a junction, a resonant conduction path (measured as a voltage peak) is realized when the cyclotron radius is half the distance between the current and voltage electrodes, corresponding the condition $B=j\cdot 2\hbar\sqrt{\pi n}/eL$, where, $j$ is the resonant mode number (physically corresponding to the number of half circles that fit between the electrodes), $e$ is the electron charge, $B$ is the magnetic field, $L$ is the distance between the electron emitter and voltage detector, and $n$ is the charge carrier density\cite{Pablo_NaturePhysics}. In a split-gate geometry, the resonant path depends on the carrier density in each region, and can be considered separately for the three distinct scenarios, shown in Fig. 1b. i) Equal density ($nn$ or $pp$): the junction is fully transparent and there is no refraction, recovering the same resonance condition given above. ii) Same carrier type but unequal density ($pp'$ or $nn'$): positive refraction across the boundary, resulting in a deviation of the resonance condition, but with carriers still focused to the voltage probe on the same side of the sample. iii) $pn'$ (unequal electron-hole densities): negative refraction occurs and there is a change in the sign of the Lorentz force, causing the charge carriers to be focused to the voltage probe on the opposite side of the sample. The sample geometry fully determines the relation between the magnetic field, $B$, and charge densities, $n_{1}$ and $n_{2}$, of the two gated regions (analytic relations defining each of the lowest resonant modes are given in the SI). For all three cases, varying the magnetic field changes the angle of incidence ($\theta_1$) at the boundary, while varying the carrier density on the right side changes both the angle of refraction ($\theta_2$) and the cyclotron radius on the right side.  Thus, by mapping out the resonance condition for transmission between the injection and collection electrodes, we can effectively measure $\theta_2$ as a function of $\theta_1$ to directly verify Snell'€™s law for both positive and negative refraction.  

An optical micrograph and schematic cross section of a typical device measured in this study are shown in Figure 1c. Monolayer graphene was encapsulated in Boron Nitride (h-BN) and placed half across a few-layer graphite bottom gate that was previously exfoliated onto an oxidized, heavily doped Si wafer. The heterostructure was then plasma etched into a rectangular shape and side-contacted using previously described techniques \cite{Wang_2013}. Independently voltage-biasing the bottom layer graphite and doped-silicon gates allows us to realize a split gate $pn$ junction. (Fig. 1b). Since a naturally cleaved graphite edge is used, the junction is expected to be atomically smooth (see SI).

In the TMF measurement, electrons are injected at one side of the graphite gated region and collected at an electrode on the opposing side, while the voltage is measured across parallel electrodes in the Si gated region (Fig 1a). Fig. 2a shows a typical result, in which the four-terminal resistance is acquired at constant hole density in the injection region ($V_{graphite}=-1$~V corresponding to a density of 6.76 $\times$10$^{11}$ cm$^{-2}$) as a function of detection side gate voltage ($V_{Si}$) and magnetic field. For the $pp'$ configuration, both the fundamental resonance and multiple higher order resonant peaks appear.  The resonance paths can not be fit to a simple $B\sim\sqrt{n}$ dependence (see SI), with the most notable deviation a pronounced kink in the second order resonance. For positive Si gate values ($pn'$ configuration) only the lowest order resonance mode is observed, with all higher orders apparently suppressed. The resonance peak is opposite in sign compared to the $pp'$ case.  This is a direct signature of carrier focusing to the upper voltage terminal. 

A detailed simulation of electron trajectories using semi-classical Billiard model\cite{beenakker1989billiard,milovanovic2014magnetic} were performed and compared to experiment (see SI). In this model, electrons are injected from the source at randomly distributed angles, weighted by a normal distribution of standard deviation $\sigma_{inj}=\pi$/15. By following their cyclotron trajectories across the junction (junction roughness is not included in the model) the probabilities of reaching the voltage probes are calculated. Transmission across the junction is modeled assuming the electronic Snell's law and momentum filtering\cite{CheinovPRB2006,RedwanPRB2012}. Fig. 2b shows the difference in probability between the two voltage leads from our simulation using identical conditions as the experiment data in Fig. 2a. The simulation matches well with the general features of our experimental data for both $pp'$ and $pn'$ cases, reproducing the trajectory of all higher order resonances in the $pp'$ condition, as well as the existence of only a single mode, with opposite sign for the $pn'$ case. Simulation reveals that the kink in $pp'$ case results from electron hitting the edge of device at the junction (see SI). For $pn'$ only lowest order is observed as the number of electrons reaching the upper electrode reduces exponentially due to filtering effect every time electrons cross $pn$ junction\cite{rickhaus2015snake,taychatanapat2015conductance}.

In both the experimental and simulated data sets, the trajectory of the lowest order resonance is well captured by our geometric model (dashed lines in Fig. 2a and 2b).  Fig. 2d shows this in more detail.  The peak position is shown as a function of $B$ and $n_{2}$ for both $pp'$ (red circles) and $pn'$ (blue circles). Also plotted are similar data points acquired by synchronizing the gates to maintain matched carrier density, giving the trajectory of the $pp$ (green circles) and $nn$ (yellow circles) response (see SI for the magnetic focusing in the matched density regime).  The theoretical resonant peak positions calculated from the geometric model are shown as solid and dashed lines. Excellent agreement is found between the peak positions and the theoretical curves for all four cases. We note that in generating the theoretical curves we use as inputs only the sample geometry (length $L=3.9$~$\mu$m, and width $W=3.9$~$\mu$m), and the gate efficiencies as extracted from Hall effect measurements (see SI), so that effectively there are no free parameters. We have repeated this measurement with three devices of varying sizes and with various gate configurations, all giving similar results. We note that for any combination of $B$, $n_{1}$, and $n_{2}$, the device geometry dictates the intersection of the electron trajectory with the junction. For each point along the first order resonant peak in Fig. 2a, we can therefore deduce the angle between the charge carrier trajectory and the boundary normal in each region. In Fig. 2e, the corresponding values of $k_{i}\sin(\theta_{i})$ for each region are plotted. The data shows a linear relation with unity slope, confirming the expected Snell's law relation for electrons. For the case of opposite carrier type, the relation shows a negative unity slope, unambiguously confirming negative refraction.

\begin{figure}
\includegraphics[width=1\linewidth]{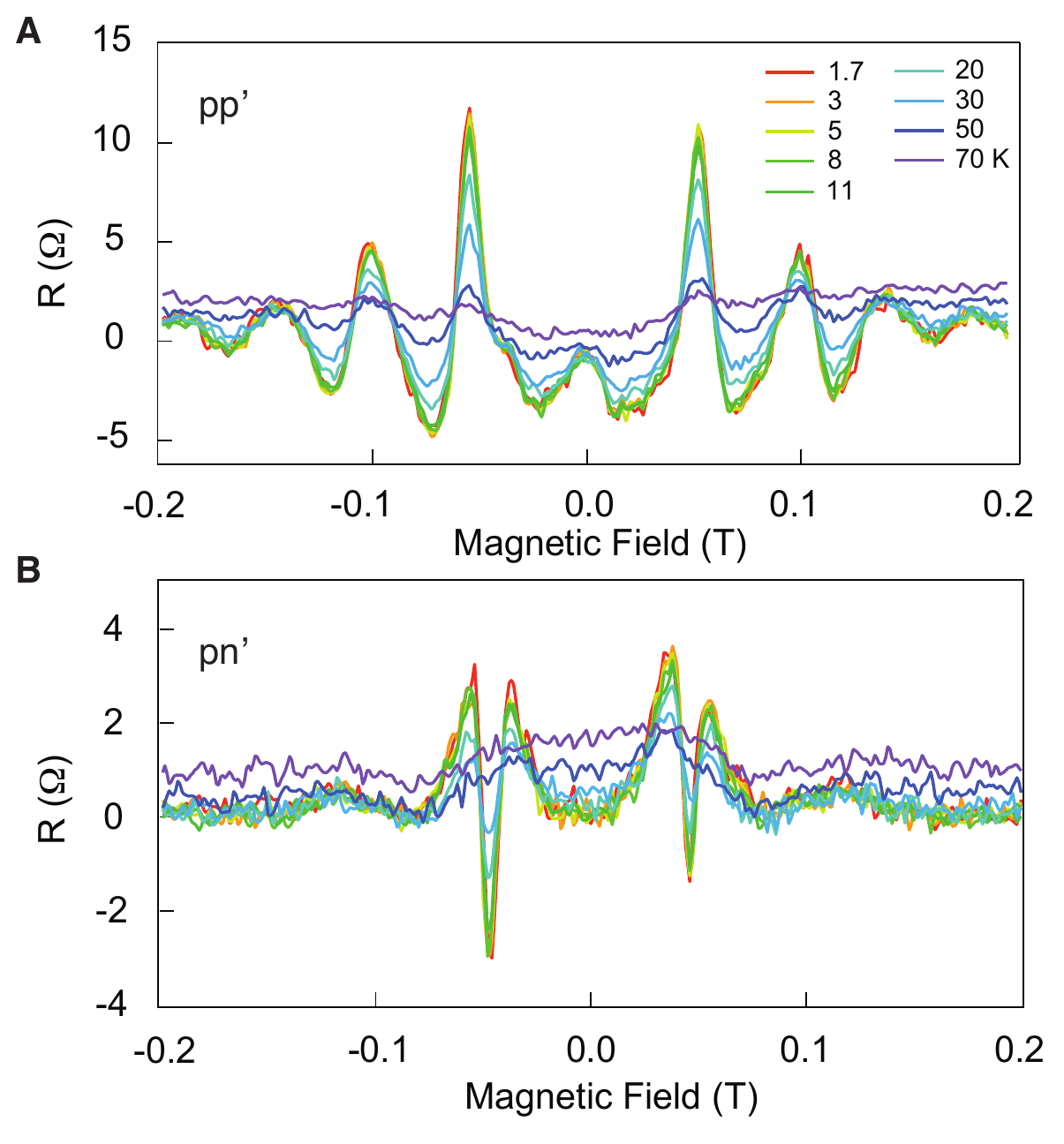}
\caption{\textbf{Temperature dependence}.(\textbf{A}) and (\textbf{B}) are temperature dependence of $pp'$ and $pn'$ resonance peaks. Data corresponds to a cut through Fig. 2a along a fixed value of $V_{Si}$.}
\label{fig:fig4}
\end{figure}

Since the points along the resonance mode can be correlated with the incidence angle, comparing the peak intensity at each point provides a measure of the angular dependent transmission coefficient across the junction.  The transmission probability across a $pn$ junction is theoretically determined by a chiral tunneling process between the bands, and depends strongly on both the incidence angle and effective junction width\cite{CheinovPRB2006,RedwanPRB2012}. For a symmetrically biased junction the transmission probability is given by\cite{CheinovPRB2006}

\begin{equation}
T \sim e^{-\pi k_F d \sin ^2 \theta}
\end{equation}

\noindent where $\theta$ is the incident and refracted angle,  $k_{F}$ is the graphene Fermi wavevector on two sides, and $d$ is the junction width. In Fig. 2f the normalized peak intensity for the $pn'$ resonance curve is plotted versus incident angle, with the blue circles and solid red line deduced from the experimental and simulated data sets, respectively. In our simulation, the transmission probability for each electron trajectory at the boundary was calculated using a more generalized form of equation 1 that allows for asymmetric bias\cite{RedwanPRB2012} (see SI). We compared experimental results with simulated response for varying junction widths (see SI), finding excellent agreement for $d=70$~nm (Fig. 2f). This is consistent with our device geometry where we anticipate a junction width on the order of $60$~nm by electrostatic modelling (see SI). Various $\sigma_{inj}$ were also tested in our simulation but no dependence was found (see SI). Our results provide strong experimental support for angle dependent transmission coefficient given by equation 1, which can be viewed as the electron equivalent of the Fresnel equations in optics, relating the transmitted and reflected probability intensities. Our findings further demonstrate that wide junctions result in selective collimation\cite{CheinovPRB2006,young2009quantum,rickhaus2013ballistic,grushina2013ballistic} of the electron beam compared to abrupt junctions with zero width (solid black line in Fig. 2f).

A striking consequence of negative refraction in graphene is Veselago lensing, in which a planar $pn'$ junction focuses diverging electrons \cite{Cheianov_Science}.  Recent transport measurement suggests evidence of this effect \cite{Gilho}, but the response is remarkably weak appearing in the signal derivative. Good agreement between our simulation and measurement for magnetic focusing, allows us to use the same model to revisit zero-field focusing across $pn'$ junctions. In Fig. 3, the transmission coefficient for our device is calculated from simulation for varying junction widths $d$. We find that, owing to the strong reflection of non-normally incident electrons, the transmission  decays rapidly with increasing $d$, and indeed, to realize transmission of 50\% compared to abrupt junction requires the $d$ to be less than 5 nm. This experimental constraint provides one explanation for why Veselago-type lensing has been difficult to achieve in previous devices and suggests scaling the $pn'$ junction width to the few nm limit to be an important criteria for realizing electron optics based on negative refraction in graphene.

Finally, owing to the interest in electron focusing for technological applications, we consider temperature dependent effects. Fig. 4a and 4b show the height of the resonant peaks as a function of magnetic field at various temperatures for  $pp'$ and $pn'$ cases, respectively. It is observed that the peak signal vanishes at around 70 K, coinciding with the temperature at which graphene mean free path becomes comparable to the resonant path length ($\sim7 ~ \mu$m) in our devices\cite{Wang_2013}. Scaling to room temperature where the graphene mean free path remains in excess of 1~$\mu$m is therefore readily feasible.

\section{\label{sec:level1}ACKNOWLEDGEMENT}
We thank P. Kim, A. Pasupathy and J.-D. Pillet for helpful discussion, and R. Ribeiro for fabrication assistance. This work is supported by the Semiconductor Research Corporation's  NRI Center for Institute for Nanoelectronics Discovery and Exploration (INDEX).

\bibliographystyle{naturemag}

\begin{thebibliography}{10}

\bibitem{Houten1995}
H.~van Houten, C.~Beenakker, {\it Confined Electrons and Photons\/} (Springer,
  1995), pp. 269--303.

\bibitem{Dragoman_ProgQuantElec1999}
D.~Dragoman, M.~Dragoman, {\it Prog. in Quantum Electronics\/} {\bf 23}, 131
  (1999).

\bibitem{Spector_APL}
J.~Spector, H.~L. Stormer, K.~W. Baldwin, L.~N. Pfeiffer, K.~W. West, {\it
  Appl. Phys. Lett.\/} {\bf 56}, 1290 (1990).

\bibitem{Shtrikman_PRB_ELens}
U.~Sivan, M.~Heiblum, C.~P. Umbach, H.~Shtrikman, {\it Phys. Rev. B\/} {\bf
  41}, 7937 (1990).

\bibitem{Cheianov_Science}
V.~V. Cheianov, V.~Fal'ko, L.~Altshuler, {\it Science\/} {\bf 315}, 1252
  (2007).

\bibitem{Park_NanoLett2008}
C.-H. Park, Y.-W. Son, L.~Yang, M.~L. Cohen, S.~G. Louie, {\it Nano Lett.\/}
  {\bf 8}, 2920 (2008).

\bibitem{Veselago_SovPhysUsp1968}
V.~G. Veselago, {\it Sov. Phys. Usp.\/} {\bf 10}, 509 (1968).

\bibitem{Shalaev_NatPhot2007}
V.~M. Shalaev, {\it Nature Photonics\/} {\bf 1}, 41 (2007).

\bibitem{Pendry_PerfectLens}
J.~B. Pendry, {\it Phys. Rev. Lett.\/} {\bf 85}, 3966 (2000).

\bibitem{shelby2001experimental}
R.~A. Shelby, D.~R. Smith, S.~Schultz, {\it science\/} {\bf 292}, 77 (2001).

\bibitem{fang2005sub}
N.~Fang, H.~Lee, C.~Sun, X.~Zhang, {\it Science\/} {\bf 308}, 534 (2005).

\bibitem{schurig2006metamaterial}
D.~Schurig, {\it et~al.\/}, {\it Science\/} {\bf 314}, 977 (2006).

\bibitem{Internal_total_reflection}
J.~R. Williams, T.~Low, M.~S. Lundstrom, C.~M. Marcus, {\it Nature
  Nanotechnology\/} {\bf 6}, 222 (2011).

\bibitem{rickhaus2013ballistic}
P.~Rickhaus, {\it et~al.\/}, {\it Nature communications\/} {\bf 4}, 2342
  (2013).

\bibitem{taychatanapat2015conductance}
T.~Taychatanapat, {\it et~al.\/}, {\it Nature communications\/} {\bf 6}, 6093
  (2015).

\bibitem{Dean}
C.~Dean, {\it et~al.\/}, {\it Nature Nanotechnology\/} {\bf 5}, 722 (2010).

\bibitem{CheinovPRB2006}
V.~V. Cheianov, V.~I. Fal'ko, {\it Phys. Rev. B\/} {\bf 74}, 041403 (2006).

\bibitem{huard2007transport}
B.~Huard, {\it et~al.\/}, {\it Physical Review Letters\/} {\bf 98}, 236803
  (2007).

\bibitem{gorbachev2008conductance}
R.~V. Gorbachev, A.~S. Mayorov, A.~K. Savchenko, D.~W. Horsell, F.~Guinea, {\it
  Nano letters\/} {\bf 8}, 1995 (2008).

\bibitem{young2009quantum}
A.~F. Young, P.~Kim, {\it Nature Physics\/} {\bf 5}, 222 (2009).

\bibitem{stander2009evidence}
N.~Stander, B.~Huard, D.~Goldhaber-Gordon, {\it Physical Review Letters\/} {\bf
  102}, 026807 (2009).

\bibitem{RedwanPRB2012}
R.~N. Sajjad, S.~Sutar, J.~U. Lee, A.~W. Ghosh, {\it Phys. Rev. B\/} {\bf 86},
  155412 (2012).

\bibitem{grushina2013ballistic}
A.~L. Grushina, D.-K. Ki, A.~F. Morpurgo, {\it Applied Physics Letters\/} {\bf
  102}, 223102 (2013).

\bibitem{rickhaus2015snake}
P.~Rickhaus, {\it et~al.\/}, {\it Nature communications\/} {\bf 6}, 6470
  (2015).

\bibitem{Topinka_Nature_2001}
M.~A. Topinka, {\it et~al.\/}, {\it Nature\/} {\bf 410}, 183 (2001).

\bibitem{Westervelt_NatPhys_2007}
K.~E. Aidala, {\it et~al.\/}, {\it Nature Physics\/} {\bf 3}, 464 (2007).

\bibitem{BaringhausSciReport2015}
J.~Baringhaus, A.~St{\"o}hr, S.~Forti, U.~Starke, C.~Tegenkamp, {\it Scientific
  reports\/} {\bf 5}, 9955 (2015).

\bibitem{Pablo_NaturePhysics}
T.~Taychatanapat, K.~Watanabe, T.~Taniguchi, P.~Jarillo-Herrero, {\it Nat.
  Phys.\/} {\bf 9}, 225 (2013).

\bibitem{Wang_2013}
L.~Wang, {\it et~al.\/}, {\it Science\/} {\bf 342}, 614 (2013).

\bibitem{beenakker1989billiard}
C.~Beenakker, H.~Van~Houten, {\it Physical review letters\/} {\bf 63}, 1857
  (1989).

\bibitem{milovanovic2014magnetic}
S.~Milovanovi{\'c}, M.~R. Masir, F.~Peeters, {\it Journal of Applied Physics\/}
  {\bf 115}, 043719 (2014).

\bibitem{Gilho}
G.-H. Lee, G.-H. Park, H.-J. Lee, {\it Nature Physics\/} {\bf 11}, 925 (2015).

\bibitem{sajjad2013manipulating}
R.~N. Sajjad, A.~W. Ghosh, {\it ACS nano\/} {\bf 7}, 9808 (2013).

\end{thebibliography}

\end{document}